\documentclass[%
 reprint,
 amsmath,amssymb,
 aps,
]{revtex4-2}

\usepackage{graphicx}% Include figure files
\usepackage{dcolumn}% Align table columns on decimal point
\usepackage{bm}% bold math
\usepackage{physics}
\usepackage{algorithm}% http://ctan.org/pkg/algorithms
\usepackage{algpseudocode}
\usepackage{float}
\newcommand{\citeasnoun}[1]{Ref.~\onlinecite{#1}}
\usepackage{xcolor}

\newcommand\NoDo{\renewcommand\algorithmicdo{}}

\begin{document}

\preprint{APS/123-QED}

\title{Efficient perturbative framework for coupling of radiative and guided modes\\
in nearly periodic surfaces}% Force line breaks with \\

\author{Sophie Fisher}
\affiliation{Department of Electrical Engineering and Computer Science, Massachusetts Institute of Technology, Cambridge, Massachusetts
02139, USA}
\email{sefisher@mit.edu}

\author{Rapha\"el Pestourie}
\affiliation{Department of Mathematics, Massachusetts Institute of Technology, Cambridge, Massachusetts
02139, USA}

\author{Steven G.~Johnson}
\affiliation{Department of Mathematics, Massachusetts Institute of Technology, Cambridge, Massachusetts
02139, USA}

\date{\today}% It is always \today, today,
             %  but any date may be explicitly specified

\begin{abstract}
We present a semi-analytical framework for computing the coupling of radiative and guided waves in slowly varying (nearly uniform or nearly periodic) surfaces, which is especially relevant to the exploitation of nonlocal effects in large-area metasurfaces. Our framework bridges a gap in the theory of slowly varying surfaces: aside from brute-force numerical simulations, current approximate methods can model \emph{either} guided \emph{or} radiative waves, but cannot easily model their \emph{coupling}. We solve this problem by combining two methods: the locally periodic approximation, which approximates radiative scattering by composing a set of periodic scattering problems, and spatial coupled-wave theory, which allows the perturbative modeling of guided waves using an eigenmode expansion. We derive our framework for both nearly uniform and nearly periodic surfaces, and we validate each case against brute-force finite-difference time-domain simulations, which show increasing agreement as the surface varies more slowly.
\end{abstract}

%\keywords{Suggested keywords}%Use showkeys class option if keyword
                              %display desired
\maketitle

%\tableofcontents

\section{Introduction}

Although many perturbative techniques are available to study \emph{guided} modes propagating \emph{through} slowly varying (nearly uniform or nearly periodic) media~\cite{Marcuse12, Katsenelenbaum98, Snyder83, PhysRevE.66.066608}, and conversely a number of approximations have been devised for \emph{radiative} waves scattering \emph{off} of slowly varying surfaces~\cite{doi:10.1021/nl302516v, Arbabi2017, 7588110, doi:10.1126/science.aaa2494, doi:10.1021/acs.nanolett.5b01727, doi:10.1021/acs.nanolett.6b05137, Arbabi:17, Su:18, Groever:18, Pestourie:18,  Perez-Arancibia:18, Verslegers:10, Cheng2017, born_wolf_bhatia_clemmow_gabor_stokes_taylor_wayman_wilcock_1999, OShea03, Voronovich94, Yu2014}, the \emph{coupling} of guided and radiative modes by slowly varying structures is relatively unstudied except by brute-force numerical simulations~\cite{LinCa21, doi:10.1021/acsphotonics.0c01058, Christiansen:20}.  This problem is especially relevant to large-area optical metasurfaces (comprising thin aperiodic subwavelength-scale patterns),  which have been widely used for free-space wavefront engineering~\cite{chen2020flat}: the large diameter of metasurface devices (often $> 1000$ wavelengths $\lambda$) can make them impractical to design without mathematical approximations~\cite{1236082, 6230714, HOLLOWAY2009100, Tretyakov15, 6891256, PhysRevB.90.235127, Jahani2016, 6415977, doi:10.1126/science.aam8100}. Such surfaces are often modeled using a locally periodic approximation (LPA), in which the far-field scattering of an incident wave from each unit cell is computed assuming a periodic surface, which works well when the unit cells are slowly varying~\cite{doi:10.1021/nl302516v, Arbabi2017, 7588110, doi:10.1126/science.aaa2494, doi:10.1021/acs.nanolett.5b01727, doi:10.1021/acs.nanolett.6b05137, Arbabi:17, Su:18, Groever:18, Pestourie:18,  Perez-Arancibia:18, Verslegers:10, Cheng2017, Yu2014}. A simplified limiting case of LPA is the locally \emph{uniform} approximation (LUA), which corresponds to LPA in the limit as the period goes to zero: LUA approximates the structure at each point by a flat surface with a matching cross section, and is also found in the form of scalar-diffraction theory~\cite{OShea03} or the Kirchhoff tangent-plane approximation (for curved surfaces)~\cite{Voronovich94}. However, LPA and LUA cannot be applied to the ``nonlocal'' problem of radiation coupling to and from guided modes since in a periodic surface guided modes do not radiate by definition~\cite{Joannopoulos08}. This is in contrast to previous ``nonlocal'' metasurfaces employing \emph{leaky} resonances~\cite{BenzaouiaJo21-QNMT,7508983, Faenzi2019, 7509680, 8486958, PhysRevB.97.115447, 8509640, BenzaouiaJo22} which radiate even for purely periodic surfaces~\cite{7105362}. On the other hand, guided-wave propagation can be treated by spatial coupled-wave theories (CWT)---eigenmode-expansion methods resulting in a set of coupled ordinary differential equations in the mode coefficients, which can be solved perturbatively for slowly varying media~\cite{Marcuse12, Katsenelenbaum98, Snyder83, PhysRevE.66.066608}. However, radiative modes are difficult to treat with CWT techniques because there is a continuous spectrum of radiative solutions, in contrast to a discrete, easily truncatable guided-mode basis.   As a  consequence, large-area metasurface designs have thus far not exploited the potential for nonlocal effects of guided modes, e.g.~for long-lifetime light-trapping to enhance light--matter interactions, metasurface lasers, or large-area grating couplers.

\begin{figure*}
    \centering
    \includegraphics[width=2\columnwidth]{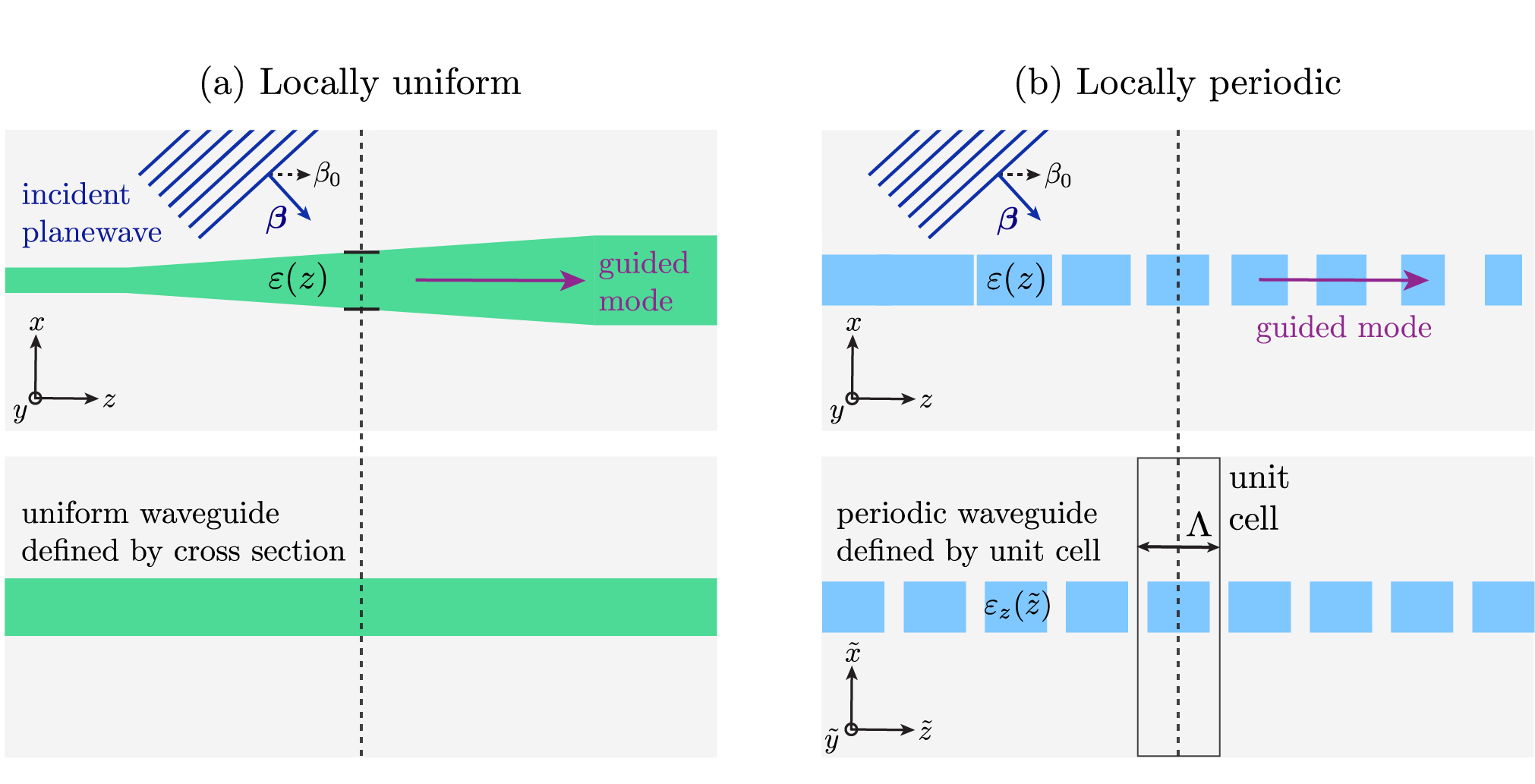}
    \caption{(\textbf{a}) An incident plane wave with propagation constant $\beta_0$ impinges on a surface with locally uniform permittivity defined by $\varepsilon(z)$. The plane wave couples to guided waves propagating along the $z$-direction. The guided waves are computed using the locally uniform approximation (LUA) and coupled-wave theory (CWT), techniques that employ a basis of completely uniform waveguides defined by the cross sections of the locally uniform surface. (\textbf{b}) The same scattering problem as in (\textbf{a}), but the surface has locally \textit{periodic} permittivity. The guided waves are computed using the locally periodic approximation (LPA) and a generalized CWT for nearly periodic media. These employ a basis of \textit{periodic} waveguides for each cross section defined by $\varepsilon_z(\tilde{z})$, with periods $\Lambda(z)$. A new coordinate, $\tilde{z}$ is introduced to describe the unit cells of the periodic waveguides. }
    \label{intro_figure}
\end{figure*}

In this paper, we bridge the gap between radiation and guided modes in slowly varying (both locally periodic and locally uniform) media by a new semi-analytical framework combining \emph{both} LPA or LUA and CWT, using LPA or LUA to model the incident or scattered radiation and using perturbatively coupled CWT to model the guided waves.
We begin by deriving and validating our framework (against brute-force numerics) for the easier case of locally uniform metasurfaces (Fig.~\ref{intro_figure}a and Sec.~\ref{locallyuniformmain}) where the cross section of the surface varies slowly with $z$.  We then generalize and validate our framework for the more complicated case of locally periodic metasurfaces (Fig.~\ref{intro_figure}b and Sec.~\ref{locallyperiodicmain}) where the unit cell varies slowly with $z$, first for the case of unit cells with a fixed period and later for a $z$-dependent period (Appendix \ref{zdependent_period}). Unlike brute-force Maxwell solvers, which require enormous resources for large-area metasurface modeling and optimization, our framework is computationally cheap: to calculate the coupling between an incident wave and guided mode, one computes the LPA solution and the guided mode fields for each unit cell (periodic Maxwell solves), and then one simply evaluates an inexpensive integral of the slowly varying overlap between the LPA solution and guided mode fields (Algorithm \ref{algorithm1}).  Compared to brute-force finite-difference time-domain (FDTD) simulations, our framework shows increasing accuracy for more slowly varying structures (where FDTD becomes much more expensive), and also exhibits high accuracy even for variation on a few-wavelength scale (Figs.~\ref{validation_uniform_figure} and \ref{validation_periodic_figure}). 

While brute-force numerical simulations capture effects due to both guided and radiation modes, optical metasurfaces are challenging to model in this way due to their irregular subwavelength-scale features and large diameters~\cite{chen2020flat,1236082, 6230714, HOLLOWAY2009100, Tretyakov15, 6891256, PhysRevB.90.235127, Jahani2016, 6415977, doi:10.1126/science.aam8100}. This is particularly the case for metasurface design by large-scale optimization (i.e.~inverse design), which requires surfaces to be simulated many times as the geometric parameters evolve~\cite{Pestourie:18, Lin:19, PhysRevApplied.9.044030, JiangFan+2020+1059+1069, Molesky2018, Chung:20, Cheng2017, bayati2020inverse, bayati2021inverse, li2021inverse}. Ways to extend the power of brute-force Maxwell solvers to the scale of metasurfaces have been proposed, e.g. using GPU-accelerated FDTD~\cite{Hughes:21, doi:10.1063/5.0071245} or spatially truncated integral-equation methods (for non-touching meta-atoms)~\cite{skarda2021simulation}. However, these methods still face severe challenges in reaching the $10^4$-wavelength scale required for inverse design of large metasurfaces. Moreover, even when computational power reaches this scale, there will always be a place for rapid-prototyping approximations that take only a few minutes on a laptop. In fact, our framework requires the periodic Maxwell solves to be performed only \emph{once} for a sequence of unit cells, and then they can be re-used (interpolated) to optimize or explore many different slowly varying metasurfaces~\cite{PhysRevE.66.066608, doi:10.1080/03052150802576797, Oskooi:12, Pestourie2020, Povinelli:05, Felici_2001}.

Many approximate techniques to model metasurfaces have already been developed. One such technique is LPA, which composes a set of periodic scattering problems for each unit cell and has been used for optimization-based inverse design~\cite{Pestourie:18, Cheng2017, bayati2020inverse, bayati2021inverse, li2021inverse} as well as selecting the phase profile of the surface \emph{a priori}~\cite{doi:10.1021/nl302516v, Arbabi2017, 7588110, doi:10.1126/science.aaa2494, doi:10.1021/acs.nanolett.5b01727, doi:10.1021/acs.nanolett.6b05137, Arbabi:17, Groever:18, Verslegers:10, Yu2014}. However, as mentioned above, LPA does not calculate any radiative coupling to and from guided modes by construction. \citeasnoun{Perez-Arancibia:18} derived a convergent series of perturbative corrections to LUA, the limit of LPA as the period goes to zero, and used higher-order terms to compute coupling to guided modes in nearly uniform structures. However, that work has not been extended to handle locally periodic surfaces, requires the implementation of complicated integral-equation operators (far more expensive than the overlap integrals in this work), and does not exploit an explicit decomposition of the fields into guided and radiative modes. Other methods of domain decomposition besides LUA/LPA have been proposed to handle strong near-field inter-cell coupling, such as overlapping-domains that model the overlapping regions from neighboring unit cells~\cite{Lin:19overlapping}, combined with absorbing-wall domains that model each unit cell with perfectly matched layer boundary conditions~\cite{Phan2019}; however, these methods explicitly discard long-range interactions and hence omit guided-mode propagation. We also note that many techniques have been developed to design large-scale metasurface antennas, filters, and other devices at optical and microwave frequencies, but these make use of \textit{leaky} waves rather than completely \textit{guided} waves and thus solve a fundamentally different problem \cite{BenzaouiaJo21-QNMT,7508983, Faenzi2019, 7509680, 8486958, PhysRevB.97.115447, 8509640, 7105362, BenzaouiaJo22}. 

Our framework builds on previous work that derived a CWT for locally periodic surfaces using a continuously varying basis of Bloch modes, derived by ``lifting'' into a higher-dimensional space of phase-shifted surfaces and projecting back down to the physical result at the end~\cite{PhysRevE.66.066608}. The end result was a set of coupled ordinary differential equations in the mode coefficients that is straightforward to apply: the coupling coefficients are given by simple modal-overlap integrals proportional to the rate of change of the unit cells, using modes computed from small \emph{periodic} Maxwell solves, and the equations can be solved to any order in the rate of change.  (One does not even need to solve for every unit cell in a long taper, but instead can interpolate the overlaps from a sample of intermediate cells~\cite{PhysRevE.66.066608, doi:10.1080/03052150802576797, Oskooi:12, Pestourie2020, Povinelli:05, Felici_2001}, achieving \emph{better} than linear scaling in the system diameter.)  However, that framework cannot be directly applied to a metasurface system in which the incident wave is a radiative mode illuminating the whole surface from the \emph{side} (as opposed to incident guided modes from the \emph{ends}), and moreover a discrete modal-expansion framework is difficult to use with a continuum of radiative modes.   To address these limitations, the key feature we introduce here is that the LPA (radiative) solution appears as a zero-th-order \emph{source} term in CWT involving an overlap integral between the LPA and guided-mode fields. As a consequence, we can solve for the guided-mode coefficients to first order in the rate of change by simple modal-overlap integrals involving the LPA source term (Algorithm \ref{algorithm1}). Our framework handles plane waves with arbitrary incident angles, which affect the LPA solution and introduce a phase velocity into the overlap integral. Moreover, one can calculate the coupling from non-plane wave sources by a plane wave expansion (Fourier transform) of the source. Our derivation and validation in this paper are for metasurfaces in two spatial dimensions that vary only in one direction $(z)$, but in the concluding remarks (Sec.~\ref{conclusion}) we discuss extension to surfaces in three spatial dimensions that vary in two directions: the key idea is that the variations in each direction decouple at first order and therefore can be treated separately. 

\section{Locally Uniform Framework}

\label{locallyuniformmain}

The problem setup is depicted in Fig.~\ref{intro_figure}a. We consider a surface and surrounding medium in two spatial dimensions that are characterized by the dielectric function $\varepsilon(x,z)$ and the magnetic permeability function $\mu(x,z)$. The medium surrounding the surface is homogeneous. We choose the surface to be locally uniform along the $z$-direction; that is,  the cross section along the $x$-direction varies continuously and slowly with $z$. A plane wave of frequency $\omega$ from the surrounding medium is incident upon the surface with wave vector $\bm{\beta} = (\beta_x, \beta_0)$ (propagation constant $\beta_0$). Here, the goal is to solve for the coupling of the incident plane wave to guided waves in the surface.

\subsection{Coupled-Wave Theory for Locally Uniform Surfaces}
To solve this problem, we apply the well-known spatial coupled-wave theory for locally uniform surfaces, following the approach and notation found in Sections II and III of~\citeasnoun{PhysRevE.66.066608}. This section serves as a review of coupled-wave theory, for which more details can be found in~\citeasnoun{PhysRevE.66.066608}.

By isolating the longitudinal $(z)$ from the transverse $(xy)$ derivatives, the fully vectorial source-free Maxwell's equations can be rewritten at a fixed frequency $\omega$ as
\begin{equation}
    \hat{A} \ket{\psi} = -i \frac{\partial}{\partial z} \hat{B} \ket{\psi} \, ,
    \label{locallyuniform_maxwelleq}
\end{equation}
where $\ket{\psi}$ is the four-component column vector
\begin{equation}
    \ket{\psi} \equiv 
    \begin{pmatrix}
    \mathbf{E}_{xy}(x,y,z)\\
    \mathbf{H}_{xy}(x,y,z)
    \end{pmatrix} e^{-i \omega t}
\end{equation}
involving the transverse ($xy$) electric and magnetic fields $\big\{ \mathbf{E}_{xy}, \mathbf{H}_{xy} \big\}$, and $\hat{A}$ and $\hat{B}$ are the matrices
\begin{equation}
    \hat{A} \equiv 
    \begin{pmatrix}
    \omega \varepsilon/c - \frac{c}{\omega} \nabla_{xy} \times \frac{1}{\mu} \nabla_{xy} \times & 0 \\[12pt]
    0 & \omega \mu/c - \frac{c}{\omega} \nabla_{xy} \times \frac{1}{\varepsilon} \nabla_{xy} \times

    \end{pmatrix},
    \label{define_Ahat}
\end{equation}

\begin{equation}
     \hat{B} \equiv 
     \begin{pmatrix}
    0 & -\mathbf{\hat{z}} \times \\
    \mathbf{\hat{z}} \times &0
    \end{pmatrix} = 
    \begin{pmatrix} & & & 1 \\
    & & -1 &  \\
    & -1 & & \\
    1 & & & 
    \end{pmatrix},
\end{equation}
where $\nabla_{xy} = \frac{\partial}{\partial x} \mathbf{\hat{x}} + \frac{\partial}{\partial y} \mathbf{\hat{y}}$. The inner product of two column vectors $\ket{\psi}$ and $\ket{\psi'}$ is given by
\begin{equation}
    \braket{\psi}{\psi'} \equiv \int \mathbf{E}_{xy}^* \cdot \mathbf{E'}_{xy} + \mathbf{H}_{xy}^* \cdot \mathbf{H'}_{xy}
    \label{innerproduct} \, ,
\end{equation}
where the integral is over the cross section along $x$ at a fixed $z$. Under this inner product, $\hat{A}$ and $\hat{B}$ are Hermitian operators for real and lossless $\varepsilon$ and $\mu$.

Eq.~\eqref{locallyuniform_maxwelleq} can be solved using spatial coupled wave theory, which employs an expansion basis of ``instantaneous" uniform waveguides defined by the cross sections of the surface (see Fig.~\ref{intro_figure}a). The fields at each cross section are expanded in terms of the instantaneous waveguide modes, resulting in a set of coupled differential equations in the mode coefficients. The instantaneous waveguide modes are found by solving Maxwell's equations [Eq.~\eqref{locallyuniform_maxwelleq}] for an infinite uniform waveguide. Since there is translational symmetry in the $z$ direction, $\ket{\psi}$ can be chosen to satisfy $\ket{\psi} = \ket{\beta} e^{i \beta z}$, where $\beta$ is the propagation constant and $\ket{\beta}$ is a $z$-independent function. Eq.~\eqref{locallyuniform_maxwelleq} becomes
\begin{equation}
    \hat{A} \ket{\beta} = \beta \hat{B} \ket{\beta}.
    \label{uniformwaveguide_eigenprob}
\end{equation}
Assuming real $\varepsilon$ and $\mu$, \eqref{uniformwaveguide_eigenprob} is a generalized Hermitian eigenproblem with eigenvalue $\beta$. It follows that there is an orthogonality relation between the eigenstates: $\bra{\beta}\hat{B}\ket{\beta'} = 0$ when $\beta \neq \beta'$, where it is assumed that $\beta$ and $\beta'$ are real (corresponding to propagating modes). If the eigenstates are guided modes, one can label the modes by an integer $n=0,1,2 \dots$ since the eigenvalues are discrete, taking the state with eigenvalue $\beta_n$ to be $\ket{n}$. The labeling can be extended to the continuum of nonguided modes as well, since in practice the eigenvalues of such modes become discrete in a finite-size computational cell. Then, the propagating (real-$\beta$) modes can be normalized such that
\begin{equation}
    \bra{m}\hat{B}\ket{n} = \delta_{m,n}\eta_n,
    \label{mode_orthonormality}
\end{equation}
where $\eta_n = \pm 1$ for forward and backward propagating modes. In coupled wave theory, Eq.~\eqref{uniformwaveguide_eigenprob} must be solved for \textit{each} cross section $(z)$ of the locally uniform surface. The eigenproblem is therefore labeled with $z$:
\begin{equation}
    \hat{A} \ket{n}_z = \beta_n(z) \hat{B} \ket{n}_z
\end{equation}
where $\ket{n}_z$ and $\beta_n(z)$ are the instantaneous eigenmodes and eigenvalues, and the $z$ dependence of $\hat{A}$ is implicit. 

The fields of the locally uniform surface are then expanded in terms of the instantaneous eigenmodes by considering the following ansatz for $\ket{\psi}$ in Eq.~\eqref{locallyuniform_maxwelleq}:
\begin{equation}
    \ket{\psi(z)} = \sum_n c_n(z) \ket{n}_z \exp \bigg(i \int^z \beta_n(z') \ dz' \bigg).
\end{equation}
where the $c_n(z)$ are eigenmode coefficients describing the inter-modal scattering along $z$, and the phase of each mode is given by an integral over $\beta_n(z)$. Next, one substitutes the ansatz into Eq.~\eqref{locallyuniform_maxwelleq} and solves for the mode evolution in $z$. The result is a linear differential equation in the mode coefficients
\begin{multline}
    \frac{d c_m}{dz} = -\eta_m \sum_{n\neq m} \frac{\bra{m}  \frac{\partial \hat{A}}{\partial z } \ket{n}_z}{\beta_n(z) - \beta_m(z)} \\ 
     \times \exp \bigg( i \int^z [\beta_n(z') - \beta_m(z')] dz' \bigg)c_n \ \\
     - \eta_m \bra{ m} \hat{B} \frac{\partial \ket{m}_z}{\partial z} c_m
\label{coupledmode_eq}
\end{multline}
that relates the mode evolution to the rate of change of the eigenoperator $\partial \hat{A} / \partial z$, and the eigenmode fields and propagation constants. Note that the right hand side of Eq.~\eqref{coupledmode_eq} includes a ``self-interaction" term given by $-\eta_m\bra{m} \hat{B} \frac{\partial \ket{m}_z}{\partial z}  c_m$. Appendix~B of~\citeasnoun{PhysRevE.66.066608} shows that by imposing a simple phase choice on the $\ket{m}_z$, corresponding to a Berry phase~\cite{sakurai_napolitano_2017}, this term can be \emph{exactly} eliminated. If the instantaneous eigenmodes are chosen to be purely real, the condition reduces to choosing the overall sign of the modes consistently. Moving forward, we will thus omit the self-interaction term. Equations~\eqref{coupledmode_eq} are the ``coupled-wave equations'' of CWT, and have been derived in a variety of notations by many authors~\cite{Marcuse12, Katsenelenbaum98, Snyder83, PhysRevE.66.066608}; its historical roots trace back to the ``telegrapher's equations'' for transmission lines.

 Eq.~\eqref{coupledmode_eq} can be solved approximately by perturbatively expanding in the rate of change of the surface cross section. One assumes that $\partial \hat{A} / \partial z$ is small, i.e.\ the surface cross section varies slowly, and expands the mode coefficients in powers of $\partial \hat{A} / \partial z$:
\begin{equation}
    c_m(z) = c_m^{(0)}(z) + c_m^{(1)}(z) + c_m^{(2)}(z) + \cdots.
    \label{powerseries}
\end{equation}
Substituting the expansion into Eq.~\eqref{coupledmode_eq} and solving to first order in $\partial \hat{A} / \partial z$ yields
\begin{multline}
    \frac{d c_m^{(1)}}{dz} = -\eta_m \sum_{n\neq m} \frac{\bra{m}  \frac{\partial \hat{A}}{\partial z } \ket{n}_z}{\beta_n(z) - \beta_m(z)} \\
     \times \exp \bigg( i \int^z [\beta_n(z') - \beta_m(z')] dz' \bigg)c_n^{(0)} \, ,
\label{firstorder}
\end{multline}
which gives the first-order mode coefficients in terms of those at zero-th order. 

\subsection{Combining CWT with LUA}
\label{combining_CWT_LUA}
Physically, we can think of the zero-th order terms as those that do not result from any inter-modal scattering and instead couple directly to the incident wave. Those at first order are the result of zero-th order terms that have undergone a single scattering process from the nonzero rate of change. Let us therefore consider the zero-th order terms in the context of an incident plane wave from the side (Fig.~\ref{intro_figure}a). A zero-th order solution to the scattering problem neglects the rate of change of the surface, and hence it conserves the propagation constant of the incident plane wave. Therefore, the zero-th order modes consist only of radiation modes with propagation constant $\beta_0$ and no guided modes. We thus set $\beta_n(z) = \beta_0$ in Eq.~\eqref{firstorder}. In principle, we can calculate the guided mode couplings to first order in $\partial \hat{A}_z / \partial z$ by computing the zero-th order radiation mode fields and propagation constants (as well as the guided mode of interest). However, this procedure is not entirely convenient, as it requires normalizing radiation modes that lie in a continuous spectrum. 

To simplify the approach, we employ LUA~\cite{Perez-Arancibia:18, born_wolf_bhatia_clemmow_gabor_stokes_taylor_wayman_wilcock_1999, OShea03, Voronovich94}, which is equivalent to LPA in the limit of zero period. In LUA, the fields scattered from a locally uniform structure are approximated by a composition of scattering problems from uniform waveguides. That is, the fields at each cross section are obtained by scattering off of the completely uniform waveguide defined by the cross section (see Fig.~\ref{intro_figure}a). While LUA yields a $z$-dependent solution that captures the surface variation, it is a zero-th order approximation because the incident field interacts only with $z$-independent structures. For an incident plane wave with propagation constant $\beta_0$, the LUA solution is given by
\begin{equation}
    \ket{\psi_\mathrm{LUA}} =  \sum_{n} c_n^{(0)}(z) \ket{n}_z  
    \label{LUAfields}
\end{equation}
where conservation of the wave vector implies that $n$ sums over radiation modes of a fixed frequency \textit{and} propagation constant ($\beta_0$), and the mode coefficients $c_n^{(0)}(z)$ are zero-th order in the rate of change. For convenience, we are factoring out the phase $e^{i\beta_0 z}$; that is,~the full zero-th order fields are given by $\ket{\psi_\mathrm{LUA}}e^{i \beta_0 z}$. 

It is clear from Eq.~\eqref{LUAfields} that the LUA solution contains the zero-th order solution to coupled wave theory for an incident plane wave. After setting $\beta_n(z) = \beta_0$ in Eq.~\eqref{firstorder}, pulling out the sum over $n$ and collecting the $n$-dependent terms, we can then insert the LUA solution:
\begin{multline}
        c_m^{(1)}(L) = -\eta_m \int_0^{L} dz \frac{\bra{m} \frac{\partial \hat{A}}{\partial z } \ket{\psi_\mathrm{LUA}} }{\beta_0 - \beta_m(z)} \\  \times \exp \bigg( i \int^z [\beta_0 - \beta_m(z')] dz' \bigg) \, .
        \label{locallyuniform_finalresult}
\end{multline}
where we have integrated both sides of Eq.~\eqref{firstorder} to solve explicitly for the guided mode coefficients. We can further simplify our result by writing the inner-product integral in terms of the guided mode and LUA fields (Eq.~(17) of~\citeasnoun{PhysRevE.66.066608}). For instance, for a locally uniform surface with constant $\mu$ and $z$-varying $\varepsilon$, Eq.~\eqref{locallyuniform_finalresult} becomes
\begin{multline}
        c_m^{(1)}(L) = \frac{-\eta_m \omega}{c} \int_0^{L} dz \frac{ \exp \bigg( i \int^z [\beta_0 - \beta_m(z')] dz' \bigg) }{\beta_0 - \beta_m(z)} \\  \times \frac{d \varepsilon}{dz} \int_{z} dx \ \mathbf{E}^*_{m} \cdot \mathbf{E}_\mathrm{LUA} \, ,
\label{locallyuniform_finalfinalresult}
\end{multline}
where $\mathbf{E}_{m}$ and $\mathbf{E}_\mathrm{LUA}$ are the three-component electric fields of the guided mode and LUA solutions respectively, and the integral over the cross section includes only the region where $d\varepsilon/dz \neq 0$, i.e.,\ between the boundaries of the waveguide.

\subsection{Validation}
\label{validation_uniform}
\begin{figure*}
    \centering
    \includegraphics[width=2\columnwidth]{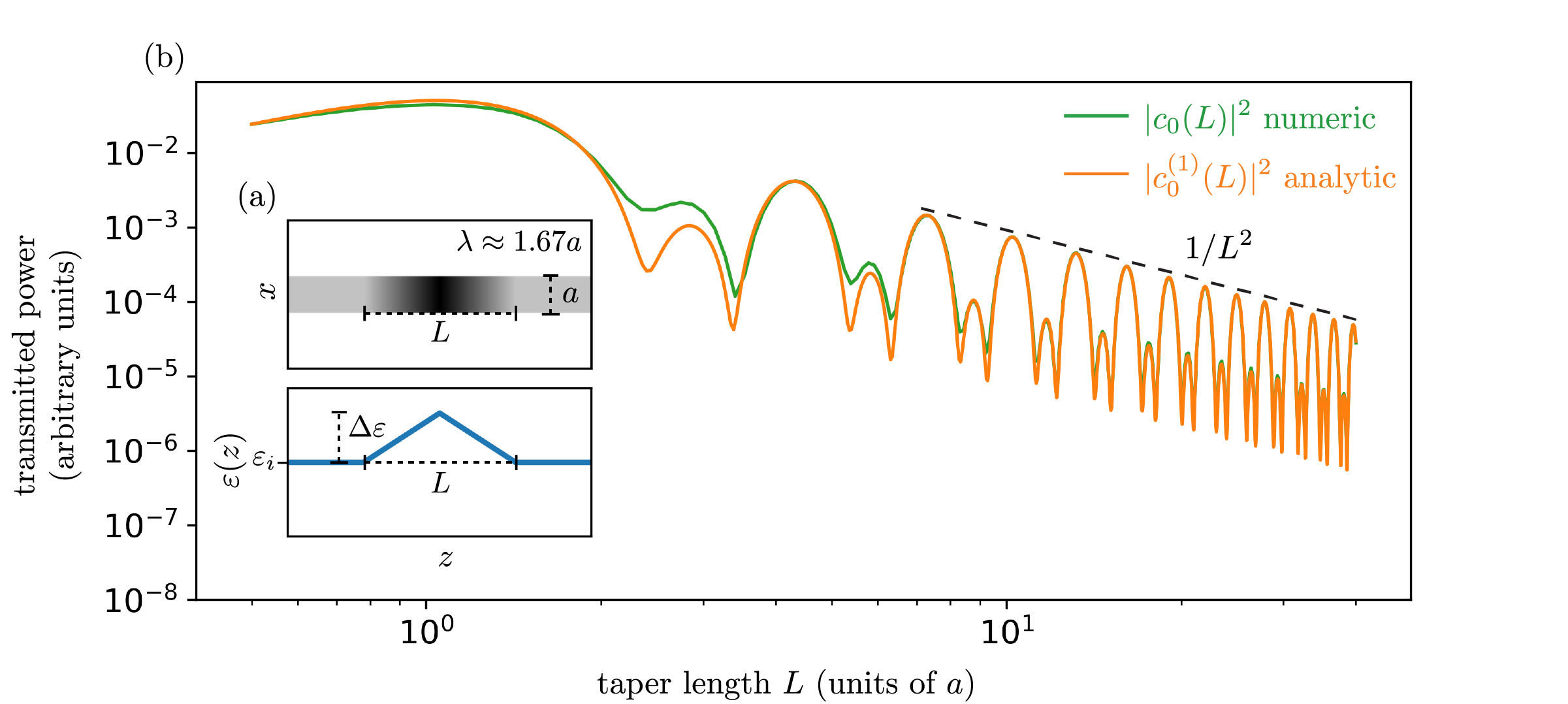}
    \caption{(\textbf{a}) Top: locally uniform surface with thickness $a$ under consideration, with permittivity indicated by the shading. Bottom: Plot of the permittivity $\varepsilon(z)$ as a function of $z$. Here $\varepsilon_i = 1.25$ and $\Delta\varepsilon = 0.5$. (\textbf{b}) Transmitted power of the fundamental TM guided mode at $\omega = 0.6 \ (2 \pi c / a)$, or $\lambda \approx 1.67~(a)$, as a function of taper length $L$. Results from brute-force FDTD simulations are shown in green, and results from our semianalytic framework [Eq.~\eqref{locallyuniform_finalfinalresult}] are shown in   orange. }
    \label{validation_uniform_figure}
\end{figure*}

Here, we validate our framework for locally uniform surfaces against brute-force FDTD simulations carried out in a freely available software package~\cite{OSKOOI2010687}. We consider a surface with locally uniform (slowly varying) permittivity $\varepsilon(z)$ depicted in Fig.~\ref{validation_uniform_figure}: a triangular taper in~$\varepsilon(z)$ joining two uniform waveguides. In this example, the uniform waveguides have permittivity $\varepsilon_i = 1.25$; the taper has an amplitude $\Delta\varepsilon = 0.5$ and a variable length $L$. The surface has a thickness $a$ and is surrounded on either side by air $(\varepsilon = 1)$. We consider a normally incident plane wave $(\beta_0 = 0)$ with TM polarization (such that $\mathbf{E} = E  \mathbf{\hat{y}}$ is perpendicular to the $xz$ plane) coupling to the fundamental TM guided mode. The operating frequency is 0.6 $(2\pi c / a)$, or $\lambda \approx 1.67~(a)$.

The choice of basis for CWT and LPA in this case is straightforward; we simply choose a set of uniform waveguides defined by the cross sections of the taper (see Fig~\ref{validation_uniform_figure}). For a particular waveguide in the basis, we compute the guided mode fields and propagation constant using a freely available eigensolver of Maxwell's equations~\cite{Johnson:01}, and we compute the LPA fields (given the same normally incident plane wave) using a one-dimensional (1D) FDTD simulation for each cross section. We then use the guided mode and LPA solutions to compute the overlap integrals over the cross sections and the phase terms in Eq.~\eqref{locallyuniform_finalfinalresult}. In principle, one needs to do this for each cross section along the taper; however, because the overlap integral and propagation constants are continuous functions of~$z$, it is sufficient to compute these quantities for only a few points and interpolate. Moreover, the interpolation can be re-used for any taper length~$L$ by rescaling with the rate of change. We note that one must be careful to choose the phase of the guided mode fields consistently across different points, which eliminates a ``self-coupling'' term from the derivation~\cite[Appendix~B]{PhysRevE.66.066608}.

Fig.~\ref{validation_uniform_figure} shows the validation results: for a range of taper lengths $L$, we compute the transmitted power of the guided mode using both our framework (orange) and a brute-force FDTD simulation (green). The plot shows excellent agreement in the large $L$ (small rate of change) limit, and relatively good agreement even for smaller $L$ comparable to the wavelength. As expected from the slope discontinuity of $\Delta\varepsilon$, the coupled power decreases as $1/L^2$, which one can show analytically by Fourier analysis via a change of variables in CWT~\cite{Oskooi:12}.

\section{Locally Periodic Framework}
\label{locallyperiodicmain}

In this section, we derive the coupling of an incident plane wave to guided waves on a locally \textit{periodic} surface. The problem setup is depicted in Fig.~\ref{intro_figure}b. The setup is the same as in Section II, except that the surface is now locally periodic in the $z$ direction rather than locally uniform; that is,~the unit cells that compose the surface vary slowly with $z$. Since a locally periodic surface becomes locally uniform in the limit of zero period, the results of this section are a strict generalization of Section~\ref{locallyuniformmain}.  As is also discussed in \citeasnoun{PhysRevE.66.066608}, the framework developed in Section~\ref{locallyuniformmain} would be a poor approximation for a photonic-crystal structure in which the periodic unit cell is rapidly varying on a subwavelength scale---it is critical to take the local periodicity into account analytically to obtain an accurate first-order approximation.

\subsection{CWT for Locally Periodic Surfaces}
To solve this problem, we apply spatial coupled-wave theory for locally \textit{periodic} surfaces, which was developed in Section IV of~\citeasnoun{PhysRevE.66.066608}. This section will largely review the coupled-wave theory developed in~\citeasnoun{PhysRevE.66.066608}; however, some modifications will be made due to the nature of the incident wave, which is then incorporated into the new LPA+CWT formulation in the following section. As in Section~\ref{locallyuniformmain}, one describes the fields scattered from the locally periodic surface by writing Maxwell's equations at a fixed frequency $\omega$ [Eqs.~(\ref{locallyuniform_maxwelleq}--\ref{innerproduct})]. However, in this case, one considers an expansion basis of eigenmodes of ``instantaneous'' \textit{periodic} waveguides at each cross section. In order to describe the waveguides and their eigenstates, one needs to introduce a ``virtual" coordinate $\tilde{z}$, to be distinguished from~$z$, that describes the variation within the unit cell of each waveguide (see Fig.~\ref{intro_figure}b). 

One may be interested in modeling a surface where the periods of the unit cells vary with $z$. If the periods of the unit cells are given by $\Lambda(z)$, one can define dimensionless coordinates scaled by $\Lambda(z)$ so that the instantaneous waveguides all have unit period. One can define $\zeta \equiv \int^z dz' / \Lambda(z')$, which effectively counts the number of periods in the locally periodic surface up to some point $z$, and $\tilde{\zeta} \equiv \tilde{z} / \Lambda(z)$. In the following, we will consider the case of a fixed period $\Lambda(z) = \Lambda$ for simplicity, in which case $\zeta= z / \Lambda$ and $\tilde{\zeta} \equiv \tilde{z} / \Lambda$. In Appendix \ref{zdependent_period}, we generalize to the case of a $z$-dependent period. For further simplicity, we assume herein that the locally periodic surface has $z$-dependent $\varepsilon$ but constant $\mu$. 

At each $z$, one defines an instantaneous periodic waveguide by $\varepsilon_z(\tilde{\zeta})$, where $\varepsilon_z(\tilde{\zeta}) = \varepsilon_z(\tilde{\zeta} + 1)$; that is,~the waveguide is unit periodic in $\tilde{\zeta}$ space. To connect the instantaneous waveguides to the physical locally periodic surface, one demands that 
\begin{equation}
    \varepsilon_z(\tilde{\zeta} = \zeta) = \varepsilon(z).
    \label{epsilon_constraint}
\end{equation}
That is, the instantaneous waveguide matches the physical surface at a single cross section, where $\tilde{\zeta} = \zeta$ (or $\tilde{z} = z$, since the period here is fixed).

Given each $\varepsilon_z(\tilde{\zeta})$, one can solve for the corresponding waveguide modes. For an infinite periodic waveguide with no current sources, Maxwell's equations at a fixed frequency $\omega$ are given by 
\begin{equation}
    \hat{A}(\tilde{\zeta}) \ket{\psi} = -\frac{i}{\Lambda} \frac{\partial}{\partial \tilde{\zeta} } \hat{B} \ket{\psi}.
    \label{periodicwaveguide_fulleq}
\end{equation}
where $\hat{A}(\tilde{\zeta})$ is $\hat{A}$ from Eq.~\eqref{define_Ahat} with $\varepsilon = \varepsilon_z(\tilde{\zeta})$ (where the $z$-dependence of the equation is suppressed for now). There is discrete translational symmetry in the $\tilde{z}/\tilde{\zeta}$ direction, so by Bloch's thoerem one can choose $\ket{\psi} = \ket{\beta} e^{i \beta \tilde{z}} = \ket{\beta} e^{i \beta \Lambda \tilde{\zeta}}$, where $\ket{\beta}$ is a unit-periodic function in  $\tilde{\zeta}$ space. Eq.~\eqref{periodicwaveguide_fulleq} becomes \begin{equation}
    \hat{C}(\tilde{\zeta}) \ket{\beta} = \beta \hat{B} \ket{\beta},
    \label{eigenproblem_periodicwaveguide}
\end{equation}
where $\hat{C}(\tilde{\zeta}) \equiv \hat{A}(\tilde{\zeta}) + \frac{i}{\Lambda} \frac{\partial}{\partial \tilde{\zeta}} \hat{B}$. As in Section \ref{locallyuniformmain}, this is a generalized Hermitian eigenproblem for real $\varepsilon$ and $\mu$, which implies an orthogonality relation between the eigenstates: $\bra{\beta}\hat{B}\ket{\beta'} = 0$ when $\beta \neq \beta'$ (assuming $\beta$ and $\beta'$ are real, corresponding to propagating and non-evanescent modes). By Bloch's theorem, eigenstates separated by reciprocal lattice vectors are equivalent up to a phase, i.e.\ $\ket{\beta + \frac{2 \pi }{\Lambda} \ell} = e^{(-2\pi i / \Lambda)\ell \tilde{z}} \ket{\beta}$ for any integer $\ell$. This implies an extended orthogonality relation:
\begin{equation}
    \bra{\beta} \hat{B} e^{(-2 \pi i\ \Lambda) \ell \tilde{z} } \ket{\beta'} = 0
    \label{extended_orthogonality}
\end{equation}
when $\beta \neq \beta' + (2\pi/\Lambda)\ell$.  Below, we choose the modes to lie within the first Brillouin zone, i.e.\ $\beta \in (-\pi/\Lambda, \pi/\Lambda]$, and label the modes $\ket{n}$ corresponding to discrete eigenvalues $\beta_n$, normalized as in Eq.~\eqref{mode_orthonormality}. Finally, since a different waveguide is defined for each cross section of the locally periodic surface, one labels the eigenequation with~$z$. Eq.~\eqref{eigenproblem_periodicwaveguide} becomes
\begin{equation}
    \hat{C}_z(\tilde{\zeta}) \ket{n(\tilde{\zeta})}_z = \beta_n(z) \hat{B} \ket{n(\tilde{\zeta})}_z
    \label{eigenequation_periodicwaveguide_zlabel}
\end{equation}
where $\hat{C}_z(\tilde{\zeta}) = \hat{A}_z(\tilde{\zeta}) + \frac{i}{\Lambda} \frac{\partial}{\partial \tilde{\zeta}} \hat{B}$, and the $\tilde{\zeta}$-dependence of the eigenstates is explicitly denoted.

To employ coupled wave theory, one needs to turn the instantaneous eigenmodes $ \ket{n(\tilde{\zeta})}_z$ into an expansion basis for $\ket{\psi(z)}$. However, the eigenmodes are defined over a unit cell, whereas one needs to expand in a single cross section. One might try to expand in the $ \ket{n(\tilde{\zeta} = \zeta)}_z$ modal cross section, but the $\tilde{\zeta}$-dependence must be retained in order to employ the orthogonality relation between the eigenmodes [Eq.~\eqref{mode_orthonormality}]. To resolve this, the approach of~\citeasnoun{PhysRevE.66.066608} is to simultaneously solve a \textit{family} of scattering problems involving different surfaces. In particular, consider the surface in $z$-space defined by $\varepsilon_z(\zeta)$. By the matching condition of Eq.~\eqref{epsilon_constraint}, $\varepsilon_z(\zeta) = \varepsilon(z)$; that is,~this defines the locally periodic surface in the physical scattering problem. Now instead, consider the entirely different surface in $z$-space defined by $\varepsilon_z(\zeta + \Delta\tilde{\zeta})$, which shifts the argument away from $\zeta$ by an amount $\Delta\tilde{\zeta}$. One can imagine constructing this surface by doing the following: at each $z$, take the periodic waveguide defined by $\varepsilon_z(\tilde{\zeta})$, and insert the cross section at $\tilde{\zeta} = \zeta + \Delta\tilde{\zeta}$. This defines an entire family of surfaces parameterized by $\Delta\tilde{\zeta}$, where $\Delta\tilde{\zeta} = 0$ corresponds to the physical surface of interest. Since the instantaneous waveguides are unit periodic in $\Delta\tilde{\zeta}$, so too are the family of surfaces; that is, $\Delta\tilde{\zeta}$ and $\Delta\tilde{\zeta} + 1$ define the same surface. Note that the notation here differs slightly from~\citeasnoun{PhysRevE.66.066608}, in that we are referring to the shift from $\zeta$ as $\Delta\tilde{\zeta}$ rather than $\tilde{\zeta}$.

In order to apply this framework to a scattering problem with an incident plane wave in free-space, our approach will differ from that of~\citeasnoun{PhysRevE.66.066608}. In particular, just as a family of different surfaces is defined, we also define different incident waves---that is, we solve a family of problems with different surfaces \textit{and} incident waves. To generate the family of incident waves, our technique will mirror that of the surfaces described above. 

Let us suppose that the electric field of the incident plane wave is given by $\vb{E}(z) = \vb{E_0} e^{i(\beta_0 z + \beta_x x)} $, where $\vb{E_0}$ is a constant three-component vector, and the $x$-dependence of $\vb{E}(z)$ is implicit. Now suppose that for each instantaneous periodic waveguide in $\tilde{z}$-space, we imagine sending in an ``instantaneous" incident plane wave with the \textit{same} wave vector. We will label the electric fields of such incident waves by $\vb{E}_z(\tilde{\zeta})$, where the $z$ subscript denotes that there is a different incident wave for \textit{each} instantaneous periodic waveguide [just as $\varepsilon_z(\tilde{\zeta})$ was defined]. In analogy to the $\varepsilon_z(\tilde{\zeta})$, we impose a constraint on the  $\vb{E}_z(\tilde{\zeta})$ to connect them to the physical incident wave:

\begin{equation}
    \vb{E}_z(\tilde{\zeta} = \zeta) = \vb{E}(z);
    \label{incident_wave_constraint}
\end{equation}
that is,~the instantaneous incident waves match the physical incident wave at the $\tilde{\zeta} = \zeta$ (or $\tilde{z} = z$) cross section. We also demand that the $\vb{E}_z(\tilde{\zeta})$ are themselves plane waves with the same wave vector as the physical incident wave. Given these constraints, there is only one possible choice of instantaneous incident waves: 
\begin{equation}
    \vb{E}_z(\tilde{\zeta}) = 
    \vb{E_0} e^{i(\beta_0 \Lambda \tilde{\zeta} + \beta_x \tilde{x})} = \vb{E_0} e^{i(\beta_0 \tilde{z} + \beta_x \tilde{x})}, 
    \label{define_instantaneous_waves}
\end{equation}
 i.e.~plane waves with the same phase as the physical plane wave. Since all of the instantaneous incident waves are the same, we no longer need the $z$ subscript, but we will keep it in order to make the analogy to the $\varepsilon_z(\tilde{\zeta})$ clear.  (For the case of variable-period structures in Appendix~A, the analogous approach leads to an incident wave that is effectively no longer a plane wave in the unit-cell problems, but which is then expanded in plane waves.)

We can now generate the family of incident waves just as~\citeasnoun{PhysRevE.66.066608} did for the family of surfaces. For instance, consider the incident wave in $z$-space defined by $\vb{E}_z(\zeta)$---by the matching condition in Eq.~\eqref{incident_wave_constraint}, this defines the incident wave in the physical scattering problem. Now, consider the incident wave in ``shifted'' $z$-space, defined by $\vb{E}_z(\zeta + \Delta\tilde{\zeta}) = \vb{E_0} e^{i \beta_0 \Lambda \Delta\tilde{\zeta}} e^{i(\beta_0 z + \beta_x x)}$. This defines a family of incident waves parameterized by $\Delta\tilde{\zeta}$ that have the same wave vector but phase-shifted amplitudes given by  $e^{i \beta_0 \Lambda \Delta\tilde{\zeta}}$. The physical incident wave of interest corresponds to $\Delta\tilde{\zeta} = 0$, i.e.\ an unshifted incident wave. 

One can now rewrite Maxwell's equations [Eq.~\eqref{locallyuniform_maxwelleq}] to solve the family of systems parameterized by $\Delta\tilde{\zeta}$:
\begin{equation}
     \hat{A}_z(\zeta + \Delta \tilde{\zeta}) \ket{\psi}_{\Delta \tilde{\zeta} }  = -i\frac{\partial}{\partial z}  \hat{B} \ket{\psi}_{\Delta \tilde{\zeta} } 
     \label{maxwelleqs_familyofsystems}
\end{equation}
where $\varepsilon_z(\zeta + \Delta\tilde{\zeta})$ has been inserted into Eq.~\eqref{locallyuniform_maxwelleq} through $\hat{A}_z(\zeta + \Delta \tilde{\zeta})$, and the fields have been labeled with $\Delta\tilde{\zeta}$. The family of incident waves described above are taken as boundary conditions. For each system, one expands $\ket{\psi}_{\Delta\tilde{\zeta}}$ in the basis of $\ket{n(\zeta + \Delta\tilde{\zeta})}_z$, i.e.\ the instantaneous waveguide modes evaluated at the $\zeta + \Delta\tilde{\zeta}$ cross section:
\begin{multline}
    \ket{\psi(z)}_{\Delta \tilde{\zeta} }  = \sum_n c_n(z, \Delta\tilde{\zeta})  \ket{n(\zeta + \Delta\tilde{\zeta}) }_z \\  \times \exp\bigg( i \int^z \beta_n(z') dz' \bigg)
    \label{ansatz_family}
\end{multline}
where the mode coefficients $c_n$ have both $z$ and $\Delta\tilde{\zeta}$-dependence. One now considers what happens to the system when $\Delta\tilde{\zeta}$ is increased by $1$, i.e.\ $\Delta\tilde{\zeta} \rightarrow \Delta\tilde{\zeta} + 1$. Since $\varepsilon_z(\tilde{\zeta})$ is unit periodic, the surface remains the same, while the incident wave remains the same up to a phase $e^{i \beta_0 \Lambda}$. Therefore the fields $\ket{\psi(z)}_{\Delta \tilde{\zeta} }$ pick up the same phase. This is in contrast to~\citeasnoun{PhysRevE.66.066608}, where the fields remained the same under $\Delta\tilde{\zeta} \rightarrow \Delta\tilde{\zeta} + 1$ (due to the incident wave being guided rather than a plane wave). Here, we choose to absorb this phase within the mode coefficients, such that $c_n(z, \Delta\tilde{\zeta} + 1) = c_n(z, \Delta\tilde{\zeta})e^{i \beta_0 \Lambda}$. Since the coefficients are therefore Bloch periodic, we can write them as the product of a periodic function and a $\Delta\tilde{\zeta}$-dependent phase:
\begin{equation}
    c_n(z, \Delta\tilde{\zeta}) = e^{i \beta_0 \Lambda \Delta\tilde{\zeta}}\tilde{c}_n(z, \Delta\tilde{\zeta})
    \label{periodic_coeff_definition}
\end{equation}
where $\tilde{c}_n(z, \Delta\tilde{\zeta} + 1) = \tilde{c}_n(z, \Delta\tilde{\zeta})$. Following~\citeasnoun{PhysRevE.66.066608}, since the $\tilde{c}_n(z, \Delta\tilde{\zeta})$ are periodic in $ \Delta\tilde{\zeta}$, one can expand the $\tilde{c}_n$ as a Fourier series:
\begin{equation}
    \tilde{c}_n(z,\Delta\tilde{\zeta}) = \sum_\ell \tilde{c}_{n,\ell}(z) e^{2 \pi i \ell \Delta\tilde{\zeta}}.
    \label{cnl_definition}
\end{equation}
Using this expression, we will formulate the coupled wave equations in the $\tilde{c}_{n,\ell}$ rather than the $\tilde{c}_n$ or $c_n$. After solving explicitly for the $\tilde{c}_{n,\ell}$, one can always recover the physical mode coefficients by the relation
\begin{equation}
    c_n(z) = \tilde{c}_n(z) = \sum_\ell \tilde{c}_{n,\ell}(z),
    \label{recover_cn}
\end{equation}
i.e.\ by setting $\Delta \tilde{\zeta} = 0$. 

Next, one substitutes the ansatz for $\ket{\psi(z)}_{\Delta\tilde{\zeta}}$ into Maxwell's equations for $\varepsilon_z(\zeta + \Delta\tilde{\zeta})$ [Eq.~\eqref{maxwelleqs_familyofsystems}]. The result is a set of differential equations in the $\tilde{c}_{n,\ell}$ that is identical to Eq.\ (25) of~\citeasnoun{PhysRevE.66.066608}: 
\begin{multline}
    \frac{d \tilde{c}_{m,k}}{dz} = -\eta_m \sum_{n,\ell\neq m,k} \frac{\bra{m}e^{2 \pi i k \tilde{\zeta}} \frac{\partial \hat{C}_z}{\partial z } e^{- 2\pi i \ell \tilde{\zeta}} \ket{n}_z}{ \Delta \beta_{n,l;m,k}(z) } \\ \times \exp \bigg( i \int^z \Delta \beta_{n,l;m,k}(z') dz' \bigg) \tilde{c}_{n,\ell} \\ -\eta^*_m \bra{m} \hat{B} \frac{ \partial \ket{m}_z}{\partial z} \tilde{c}_{m,k}
\label{coupledmodeeq_locallyperiodic}
\end{multline}
where the phase mismatch $\Delta\beta$ is given by
\begin{equation}
      \Delta \beta_{n,\ell;m,k}(z)\equiv \beta_n(z) - \beta_m(z) + \frac{2 \pi}{\Lambda}(\ell-k)  
\end{equation}
As in Section \ref{locallyuniformmain}, the ``self-interaction" term $-\eta^*_m \bra{m} \hat{B} \frac{ \partial \ket{m}_z}{\partial z} c_{m,k}$ can be \emph{exactly} eliminated by a straightforward ``Berry'' phase choice for $\ket{m}_z$~\cite[Appendix~B]{PhysRevE.66.066608}. This involves evaluating simple overlap integrals between instantaneous eigenmodes at nearby points~$z$. We thus omit this term below. One can approximately solve Eq.~\eqref{coupledmodeeq_locallyperiodic} by perturbatively expanding the mode coefficients in the rate of change. That is, one assumes that $\partial \hat{C}_z(\tilde{\zeta})/\partial z$ is small, i.e.\ the instantaneous unit cell varies slowly with $z$, and expands the $\tilde{c}_{m,k}$ in powers of $\partial \hat{C}_z(\tilde{\zeta})/\partial z$:
\begin{equation}
    \tilde{c}_{m,k}(z) = \tilde{c}_{m,k}^{(0)}(z) + \tilde{c}_{m,k}^{(1)}(z) + \tilde{c}_{m,k}^{(2)}(z) + \cdots
    \label{powerseries_locallyperiodic}
\end{equation}
Substituting the expansion into Eq.~\eqref{coupledmodeeq_locallyperiodic} and solving to first order, one finds
\begin{multline}
    \frac{d \tilde{c}_{m,k}^{(1)}}{dz} = -\eta_m \sum_{n,\ell\neq m,k} \frac{\bra{m}e^{2 \pi i k \tilde{\zeta}} \frac{\partial \hat{C}_z}{\partial z } e^{- 2\pi i \ell \tilde{\zeta}} \ket{n}_z}{ \Delta \beta_{n,l;m,k}(z) } \\ \times \exp \bigg( i \int^z \Delta \beta_{n,l;m,k}(z') dz' \bigg) \tilde{c}_{n,\ell}^{(0)}
    \label{firstorderperiodic_betan}
\end{multline}
which gives the first order mode coefficients in terms of those at zero-th order (obtained below from the LPA solution). 

\subsection{Combining CWT with LPA}
Here we can make a similar argument to that in Section \ref{locallyuniformmain} regarding the zero-th order modes. Since the zero-th order solution necessarily neglects the rate of change of the unit cell, it must conserve the wave vector of the incident plane wave up to $2\pi / \Lambda$. Therefore, the zero-th order modes are radiation modes with $\beta_n(z) = \beta_0$ (since we have arbitrarily restricted the modes to lie within the first Brillouin zone). Then Eq.~\eqref{firstorderperiodic_betan} becomes
\begin{multline}
    \frac{d \tilde{c}_{m,k}^{(1)}}{dz} = -\eta_m \sum_{n,\ell\neq m,k} \frac{\bra{m}e^{2 \pi i k \tilde{\zeta}} \frac{\partial \hat{C}_z}{\partial z } e^{- 2\pi i \ell \tilde{\zeta}} \ket{n}_z}{ \Delta \beta_{l;m,k}(z) } \\  \times \exp \bigg( i \int^z \Delta \beta_{l;m,k}(z') dz' \bigg) \tilde{c}_{n,\ell}^{(0)}
    \label{firstorder_locallyperiodic}
\end{multline}
where the phase mismatch $\Delta\beta$ is now given by 
\begin{equation}
     \Delta \beta_{l;m,k}(z)\equiv \beta_0 - \beta_m(z) + \frac{2 \pi}{\Lambda}(\ell-k). 
\end{equation}
We have found an expression for guided mode coefficients to first order in $\partial \hat{C}_z(\tilde{\zeta})/\partial z$. However, in its current form, we must solve explicitly for and normalize the zero-th order radiation modes. To avoid this, we will express Eq.~\eqref{firstorder_locallyperiodic} in terms of the fields given by LPA, which is a generalization of LUA in Section \ref{combining_CWT_LUA} to structures with nonzero periods.

In LPA, the fields scattered from a locally periodic surface are approximated by a composition of scattering problems from periodic waveguides~\cite{doi:10.1021/nl302516v, Arbabi2017, 7588110, doi:10.1126/science.aaa2494, doi:10.1021/acs.nanolett.5b01727, doi:10.1021/acs.nanolett.6b05137, Arbabi:17, Su:18, Groever:18, Pestourie:18,  Perez-Arancibia:18, Verslegers:10, Cheng2017, Yu2014}. In this case, it is natural to apply LPA alongside coupled-wave theory since we have already defined instantaneous periodic waveguides and incident waves corresponding to each cross section.

Suppose we want to apply LPA to the physical surface of interest. For a given cross section, we simply take the instantaneous waveguide defined by $\varepsilon_z(\tilde{\zeta})$ and send in the instantaneous incident wave with electric field $\vb{E}_z(\tilde{\zeta})$. For each $z$, this produces a solution that is defined over a unit period in $\tilde{\zeta}$ space. Since the incident wave is a plane wave with propagation constant $\beta_0$, conservation of the wave vector implies that we can write the solution in $\tilde{\zeta}$ space as a sum over radiation modes with fixed frequency \textit{and} propagation constant $(\beta_0)$:
\begin{equation}
    \ket{\psi_\mathrm{LPA}(\tilde{\zeta})}_z = \sum_n b_n^{(0)}(z) \ket{n(\tilde{\zeta})}_z
    \label{LPA_solution}
\end{equation}
where the radiation modes are weighted by coefficients $b_n^{(0)}(z)$ that depend on $z$ but not $\tilde{\zeta}$. As denoted by the superscript, the coefficients are zero-th order in $\partial \hat{C}_z(\tilde{\zeta})/\partial z$ because the incident waves scatter off of completely periodic structures. For convenience, we are factoring out the phase $e^{i\beta_0 z}$ from the LPA fields; i.e.\ the full fields are given by $\ket{\psi_\mathrm{LPA}(\tilde{\zeta})} e^{i \beta_0 \tilde{z} }$. Hence, when we phase-shift the unit cells by $\Delta\tilde{\zeta}$, the corresponding LPA fields for the shifted problem are $\ket{\psi_\mathrm{LPA}(\zeta + \Delta\tilde{\zeta})}_z e^{i \beta_0 \Lambda (\zeta + \Delta\tilde{\zeta}) }$.

\begin{figure*}
    \centering
    \includegraphics[width=2\columnwidth]{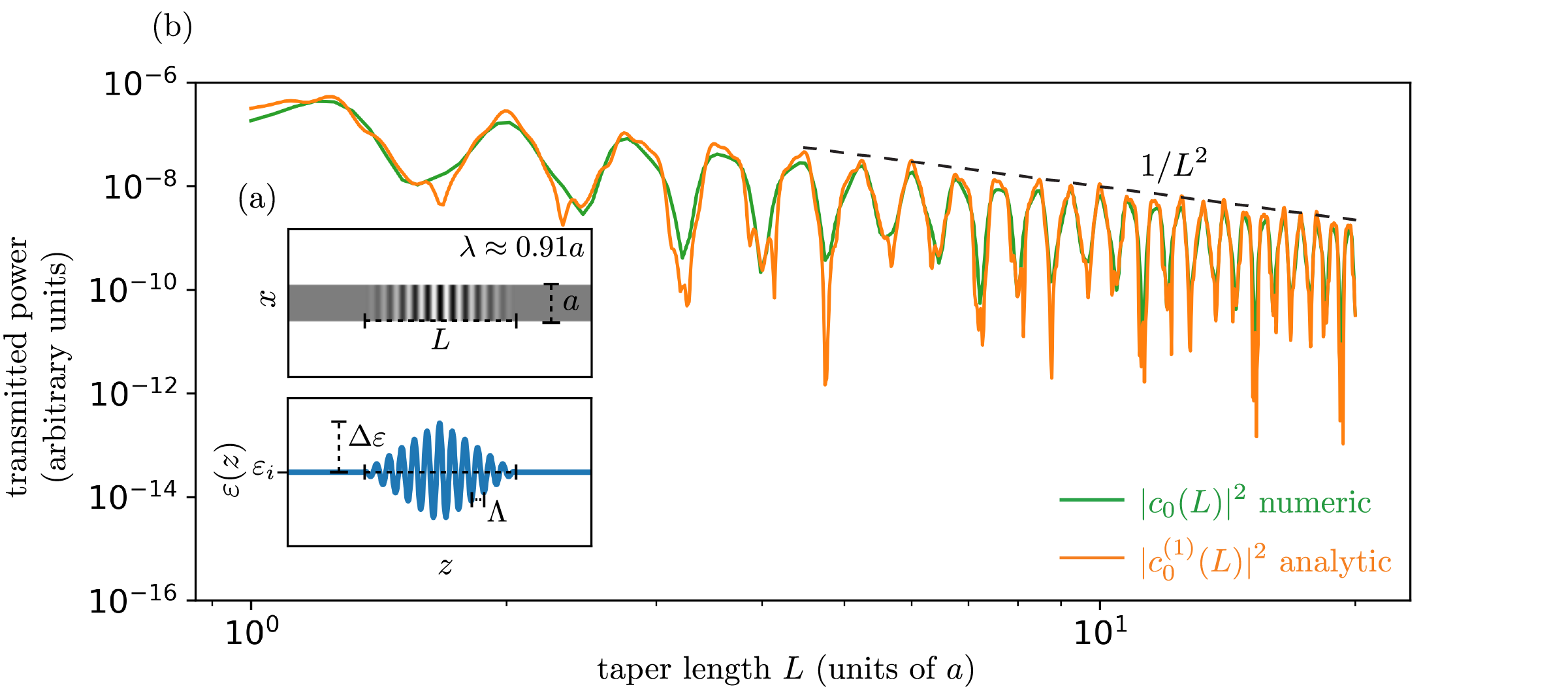}
    \caption{ (\textbf{a}) Top: locally periodic surface with thickness $a$ under consideration. Bottom: Plot of the permittivity $\varepsilon(z)$ as a function of $z$, a sine wave with fixed period and a triangular taper in the amplitude. Here $\varepsilon_i = 2$, $\Delta\varepsilon = 0.2$, and $\Lambda = 0.25~(a)$. (\textbf{b}) Transmitted power of the fundamental TM guided mode at $\omega = 1.1 \ (2 \pi c / a)$, or $\lambda \approx 0.91 ~(a)$, as a function of taper length $L$. Results from brute-force FDTD simulations are shown in green, and results from our semianalytic framework [Eq.~\eqref{locallyperiodic_finalfinalresult}] are shown in   orange. }
    \label{validation_periodic_figure}
\end{figure*}

Next, we need to understand how the $b_n^{(0)}$ in \eqref{LPA_solution} relate to the Fourier coefficients $\tilde{c}_{n,\ell}^{(0)}$ so that we can insert $\ket{\psi_\mathrm{LPA}}$ into Eq.~\eqref{firstorder_locallyperiodic}. First, if we compare the expression for the full LPA fields (including the $\Delta\tilde{\zeta}$-dependent phase) to the modal expansion of Eq.~\eqref{ansatz_family} assuming that $\beta_n(z) = \beta_0$, we find that the $b_n^{(0)}$ are equivalent to the $\tilde{c}_n^{(0)}$; that is,~they are periodic in $\Delta\tilde{\zeta}$. Moreover, by Eq.~\eqref{LPA_solution}, the $b_n^{(0)}$ are independent of $\tilde{\zeta}$ and therefore of $\Delta\tilde{\zeta}$. Hence, by Eq.~\eqref{cnl_definition}, the $b_n^{(0)}(z)$ are equivalent to the $\ell = 0$ Fourier coefficients $\tilde{c}_{n,0}^{(0)}$. Since the LPA solution is the zeroth-order solution to coupled-wave theory for an incident plane wave, we can drop the sum over $\ell$ in Eq.~\eqref{firstorder_locallyperiodic}, leaving only the $\ell =0$ term. Then, after pulling out the sum over $n$ and collecting the $n$-dependent terms, we can insert the LPA solution into Eq.~\eqref{firstorder_locallyperiodic}:
\begin{multline}
    \tilde{c}_{m}^{(1)}(L) = -\eta_m \int_0^{L} dz \sum_k  \frac{\bra{m}e^{2 \pi i k \tilde{\zeta}} \frac{\partial \hat{C}_z}{\partial z } \ket{\psi_\mathrm{LPA}}_z }{ \Delta \beta_{0;m,k}(z) } \\  \times \exp \bigg( i \int^z \Delta \beta_{0;m,k}(z') dz' \bigg) 
    \label{locallyperiodic_finalresult}
\end{multline}
where we have integrated both sides over $z$ and summed over $k$ to solve explicitly for the $\tilde{c}_{m}^{(1)}$ [Eq.~\eqref{recover_cn}]. Finally, we can simplify this result by expressing the inner-product integral in terms of the guided mode and LPA fields (Eq.\ (30) of~\citeasnoun{PhysRevE.66.066608}). Assuming that the locally periodic surface has $z$-independent $\mu$, Eq.~\eqref{locallyperiodic_finalresult} becomes
\begin{multline}
    \tilde{c}_{m}^{(1)}(L) = \frac{- \eta_m \omega}{c} \int_0^{L} dz \sum_k \frac{ \exp \bigg( i \int^z \Delta \beta_{0;m,k}(z') dz' \bigg) }{ \Delta \beta_{0;m,k}(z) } \\ \times  \int_{z} d\tilde{x} d\tilde{z} e^{(2 \pi i / \Lambda ) k \tilde{z}} \frac{\partial \varepsilon_z(\tilde{\zeta})}{\partial z} \mathbf{E}^*_{m} \cdot \mathbf{E}_\mathrm{LPA}
\label{locallyperiodic_finalfinalresult}
\end{multline}
where $\mathbf{E}_{m}$ and $\mathbf{E}_\mathrm{LPA}$ are the three-component electric fields of the guided mode and LPA solutions respectively. Eq.~\eqref{locallyperiodic_finalfinalresult} also assumes that $\partial \varepsilon_z(\tilde{\zeta}) / \partial z$ has no moving boundary---to handle such cases, the integral must be written in terms of the field components that are continuous across the boundary, $\mathbf{E}_{\parallel}$ and $D_{\perp}$~\cite[Eq.~(12)]{PhysRevE.65.066611}.

\subsection{Validation}

\begin{figure*}
\begin{minipage}{\linewidth}
\begin{algorithm}[H]
  \caption{Semi-analytical radiative/guided-mode coupling for slowly varying periodic surfaces}
  \label{algorithm1}
   \begin{algorithmic}[1]
   
   \NoDo
\For{a set of points $s \in [0,1]$ }
\State{Compute the LPA solution $\ket{\psi_{LPA}}_{s}$: periodic scattering solve }
\State Compute the phase-corrected, normalized eigenmode $\ket{m}_{s}$  and $\beta_m(s)$: periodic eigensolve 
\For{$k = -3$ to $3$ }
\State{Compute overlap integrals $M_k[s(u)]$ and phase mismatches $\Delta \beta_k[s(u)]$ (Eq.~\ref{locallyperiodic_finalresult_scaled})}
\EndFor
\EndFor
%\item[]
\State{\textbf{for} each $k$: Form interpolating polynomials or splines for $M_k[s(u)]$ and $\Delta \beta_k[s(u)]$ to obtain values at any $s$}
   
   \end{algorithmic}
\end{algorithm}
\end{minipage}
\end{figure*}

Here, we validate our framework for locally periodic surfaces against brute-force FDTD simulations. We also provide an algorithm which describes how to efficiently apply our framework to compute the coupling to guided waves in such surfaces. We consider a surface that is analogous to the example in Section \ref{validation_uniform}: the permittivity is described by a sine wave with a fixed period and a triangular taper in the amplitude, with uniform waveguides on either side of the taper. That is, the permittivity is given by $\varepsilon(z) = \varepsilon_i + \Delta\varepsilon A(z) \sin{(2 \pi z / \Lambda)}$, where $A(z)$ describes a triangular taper with unit amplitude and width $L$ (see Fig.~\ref{validation_periodic_figure}). In this case, we choose $\varepsilon_i = 2$ (larger than in Sec.~\ref{validation_uniform} in order to ensure a strongly localized guided mode), $\Delta\varepsilon = 0.2$, and $\Lambda = 0.25a$. The surface has a thickness of $a$ and is surrounded on either side by air. We consider a normally incident plane wave $(\beta_0 = 0)$ with TM polarization coupling to a fundamental TM guided mode, operating at a frequency of 1.1 $(2\pi c / a)$. 

In order to choose a basis of waveguides, we must obey the matching condition of Eq.~\eqref{epsilon_constraint}. While there are many choices that satisfy this condition, the most straightforward one is as follows: for each $z$, the permittivity is given by a sine wave (with period $\Lambda$) whose amplitude matches that of the physical taper at $z$. That is, the instantaneous waveguides are described by $\varepsilon_z(\tilde{z}) = \varepsilon_i + \Delta\varepsilon A(z) \sin{(2 \pi \tilde{z} / \Lambda)}$. As described in Section~VI of~\citeasnoun{PhysRevE.66.066608}, one must be careful to choose a taper where the guided mode remains guided throughout the taper, e.g.,~where the operating frequency does not fall into a bandgap at some intermediate points---in such cases, the theory breaks down, and the transmission of the mode becomes poor. As in Section \ref{validation_uniform}, for a particular instantaneous waveguide, we compute the guided mode and LPA solutions, then use these to compute the overlap integrals and phase terms in Eq.~\eqref{locallyperiodic_finalfinalresult}. While Eq.~\eqref{locallyperiodic_finalfinalresult} sums over infinitely many integer values of $k$, in practice one can drop most of these terms. This is because the largest contributions come from terms where $\Delta\beta_{0;m,k}(z)$ is smallest due to the integral over the phase in Eq.~\eqref{locallyperiodic_finalfinalresult}. In this case, $\Delta\beta_{0;m,k}(z)$ may be smallest when $k=-1$, $0$, or $1$ depending on the propagation directions of the incident wave and guided mode (if both propagate in the same direction, it is when $k=0$). We found that it is sufficient to sum over the five terms centered around the smallest $\Delta\beta_{0;m,k}(z)$; thus using the seven terms from $k=-3$ to $3$ should always be more than enough (In this particular example, the average relative error between five and seven terms was $0.0037\%$.) As in Section \ref{validation_uniform}, rather than computing the overlap integrals and propagation constants for each point along the taper, we do so only for a few points and interpolate the remainder — in this case, we perform separate interpolations of the overlap integral for each value of $k$. After doing so, the interpolations can be applied to any taper length $L$ by re-scaling with the rate of change. As before, one must consistently choose the phase of the guided mode fields in order to disregard the “self-interaction” term omitted in the derivation. 

Fig.~\ref{validation_periodic_figure} shows the validation results: for each taper length $L$, we compute the transmitted power of the guided mode using both our framework (orange) and an FDTD simulation (green), which show agreement in the large $L$ limit. In the following, we provide an algorithm that summarizes the steps involved in computing the coupling to a guided mode using our framework, assuming a surface whose unit cells have a fixed period, $\Lambda(z) = \Lambda$. In order to do so, we will rewrite Eq.~\eqref{locallyperiodic_finalresult} in a form that is more convenient to apply. First, we define a dimensionless coordinate $u = z/L$ such that the surface varies only from $u=0$ to $u=1$. Next, following~\citeasnoun{doi:10.1080/03052150802576797, Oskooi:12, Povinelli:05}, we define a dimensionless parameter $s \in [0,1]$ that characterizes the instantaneous unit cells of the surface. For instance, for a dielectric waveguide composed of a sequence of periodic flanges, $s$ could be proportional to the flange height (see Fig.~1 of~\citeasnoun{Oskooi:12}). Then, a given surface taper is described by the continuous function $s(u)$, which defines a sequence of instantaneous unit cells for each point of the taper. In general, the taper may be described by multiple independent parameters $s_n(u)$; however we assume only one parameter for simplicity. Then, Eq.~\eqref{locallyperiodic_finalresult} can be rewritten as 
\begin{multline}
    \tilde{c}_{m}^{(1)}(L) = -\eta_m \int_0^{1} du \ s'(u) \sum_k  \frac{ M_k[s(u)] }{ \Delta \beta_k[s(u)] } \\ \times \exp \bigg( i \int^u \Delta \beta_k[s(u')] du' \bigg) 
    \label{locallyperiodic_finalresult_scaled}
\end{multline}
where $M_k[s(u)] = \bra{m}e^{2 \pi i k \tilde{\zeta}} \frac{\partial \hat{C}_z}{\partial z } \ket{\psi_\mathrm{LPA}}_z$ and $\Delta \beta_k[s(u)] = \Delta \beta_{0;m,k}(z)$ are the $k$-dependent matrix elements and phase mismatches respectively. Using this notation, the algorithm for computing the coupling to a guided wave in a nearly periodic surface with fixed period, $\Lambda(z) = \Lambda$ is described in Algorithm \ref{algorithm1}.

Step~1 could employ a variety of interpolation algorithms, such as cubic splines or Chebyshev polynomials, to obtain high accuracy from a small number of samples. Once one has carried out the algorithm for a particular taper with length $L$ and rate profile $s(u)$, computing the coupling for additional tapers requires minimal effort. In particular, one simply redefines $s(u) \in [0,1]$ and $L$ and executes Step 8, since the calculations used to sample the interpolating points in Steps 1--7 can be re-used. This method was used in~\citeasnoun{doi:10.1080/03052150802576797, Oskooi:12} to optimize a taper profile for guided mode transmission, which requires evaluating CWT for many different $s(u)$. Finally, as described in Appendix \ref{zdependent_period}, this algorithm can applied to structures with $z$-dependent periods by Fourier transforming the incident wave in $\zeta$ and applying the algorithm to each plane wave component. 

\section{Concluding Remarks}
\label{conclusion}
We have developed a semi-analytical framework combining LPA and CWT that solves the scattering problem of incident radiation coupling to guided waves in large-area, slowly varying surfaces. Our framework is particularly relevant to optical metasurfaces, where solving this problem by brute-force simulation is formidable due to their large diameters, and where existing analytical approximations have largely focused on the problem of incident radiation coupling to outgoing radiation or leaky modes. A possible application of this work is to metasurface design through numerical optimization, building on previous optimization of metasurfaces by LPA~\cite{Pestourie:18, Cheng2017, bayati2020inverse, bayati2021inverse, li2021inverse} and guided-mode couplers using CWT~\cite{doi:10.1080/03052150802576797, Oskooi:12, Povinelli:05, Felici_2001}---such as for large-area grating couplers, for nonlinear processes and lasing via long-lifetime trapping in guided modes, and for other devices exploiting ``nonlocal'' effects due to transport by guided waves in large surfaces. 

Our derivation could be straightforwardly generalized in a variety of ways. For example, to other linear materials, such as anisotropic, magnetic, absorptive (complex~$\beta$), or even magneto-optic or chiral media.   Arbitrary incident waves can be treated by expanding them in plane waves, applying our framework to each plane wave component.  Coupling radiative modes to guided modes, as in the examples above, is exactly equivalent to coupling guided modes to radiation under a reciprocity or time-reversal transformation~\cite{Landau84}.  Surfaces with a $z$-varying periodicity can be handled by a coordinate transformation as in~\citeasnoun{PhysRevE.66.066608} (Appendix.~\ref{zdependent_period}).   Although the examples in this paper were for 1D-patterned surfaces in two dimensinos, exactly the same equations are applicable in three dimensions for structures that are nearly periodic along \emph{one} direction, such as waveguide tapers~\cite{Povinelli:05}.

A more subtle generalization would be to two-dimensional patterned surfaces in three dimensions that are nearly periodic along \emph{two} directions ($xy$), as in most practical metasurface devices. One promising approach is to perturbatively decouple the two directions: for variation in two directions, one can Taylor expand the mode coefficients in both $\partial \hat{C}_x(\tilde{\zeta})/\partial x$ and $\partial \hat{C}_y(\tilde{\zeta})/\partial y$ (the slow rates of change in $x$ and $y$). To first order in these rates, the variations in each direction decouple: the mode coefficient is given by a term that is first order in $\partial \hat{C}_x(\tilde{\zeta})/\partial x$ plus another term that is first order in $\partial \hat{C}_y(\tilde{\zeta})/\partial y$. These terms could be computed independently by applying Eq.~\eqref{locallyperiodic_finalresult} to the variation in each direction (while approximating the surface as completely periodic in the perpendicular direction, for each ``line'' of unit cells).  (A practical obstacle to research on similar semi-analytical methods in three dimensions is simply that brute-force validation becomes vastly more expensive. This can be combated by carefully choosing the three-dimensional test problem, such as an array of \emph{disconnected} high-symmetry scatterers amenable to fast multipole or integral-equation methods~\cite{skarda2022low, GIMBUTAS201322, EGEL2017103}, with appropriate absorbing-boundary techniques for the guided modes~\cite{Pissoort2003,ZhangLe11,Bruno2017}.)

Moreover, our work could serve as the starting point to compute \emph{next}-order scattering processes, such as the second-order coupling of incident to outgoing radiation involving an intermediate scattering to guided modes (e.g.,~to capture nonlinear effects experienced by the guided modes), or simply as a validation check for the first-order calculation.

\section{Acknowledgements}
This work was supported in part by the U.S. Army
Research Office through the Institute for Soldier Nanotechnologies at MIT under Award Nos.~W911NF-18-2-0048 and~W911NF-13-D-0001, and by the Simons Foundation collaboration on Extreme Wave Phenomena. R.P. is also supported in part by MIT-IBM Watson AI Laboratory (Challenge No. 2415). \appendix

\section{Extension to surfaces with $z$-dependent period $\Lambda(z)$}
\label{zdependent_period}

In Section \ref{locallyperiodicmain}, we considered only surfaces where the period of the instantaneous unit cell is $z$-independent, i.e.\ $\Lambda(z) = \Lambda$. Here, we extend our framework
to handle a $z$-dependent period, which will reduce to solving multiple problems with $z$-independent periods. Recall that, in anticipation of this case, we introduced dimensionless coordinates scaled by $\Lambda(z)$. In particular, we defined $\zeta \equiv \int^z dz' / \Lambda(z')$, which counts the number of periods in the locally periodic surface up to some point $z$, and $\tilde{\zeta} \equiv \tilde{z} / \Lambda(z)$, which scales all of the instantaneous gratings to have unit period. Using these coordinate systems and assuming a $z$-dependent period, all of the analysis carried out in Eqs.~(\ref{epsilon_constraint}--\ref{eigenequation_periodicwaveguide_zlabel}) still holds. However, a problem arises when we try to define the instantaneous incident waves [Eqs.~(\ref{incident_wave_constraint}--\ref{define_instantaneous_waves})]. That is, for general $\Lambda(z)$ we cannot define $\vb{E}_z(\tilde{\zeta})$ that are \textit{plane waves} while satisfying the constraint of Eq.~\eqref{incident_wave_constraint}. As a result, the family of incident waves parameterized by $\Delta\tilde{\zeta}$ will no longer be plane waves phase-shifted by $e^{i \beta_0 \Lambda \Delta\tilde{\zeta}}$, and in general will not be plane waves at all. Thus, the approach requires some modification.

Intuitively, the problem stems from trying to express the incident plane wave in terms of the scaled coordinates. In particular, the physical incident wave [with electric field $\vb{E}(\zeta)$] is \emph{no longer a plane wave} in $\zeta$-space, since it has effectively been re-scaled by $\Lambda(z)$ into an arbitrary incident wave. Using scaled coordinates transforms the original scattering problem with an incident plane wave and $z$-dependent period into one with an arbitrary incident wave and $z$-independent period, since the $\varepsilon_z(\tilde{\zeta})$ are all unit periodic in $\tilde{\zeta}$. Therefore, we can treat this problem using the framework that we developed for surfaces with a $z$-independent period. In particular, our solution is to Fourier transform the incident wave in $\zeta$-space into a plane wave basis and apply Eq.~\eqref{locallyperiodic_finalresult} to each plane wave component. To get the total coupling to guided modes, the couplings from each plane wave can simply be added together due to the linearity of Maxwell's equations. For instance, suppose the electric fields of the physical incident wave are given by
\begin{equation}
    \vb{E}(\zeta) = \int d \bar{\beta}_0 d \bar{\beta}_x \boldsymbol{\alpha}(\bar{\beta_0}, \bar{\beta}_x) e^{i (\bar{\beta_0} \zeta + \bar{\beta}_x x)}
\end{equation} 
where $\boldsymbol{\alpha}(\bar{\beta_0}, \bar{\beta}_x)$ is given by the Fourier transform of $\vb{E}(\zeta)$. Then, for each nonzero $\boldsymbol{\alpha}(\bar{\beta_0}, \bar{\beta}_x)$ such that $\bar{\beta}_x < 0$, i.e.~the plane wave is traveling towards the surface rather than away from it, one computes the coupling from the plane wave by applying Eq.~\eqref{locallyperiodic_finalresult} assuming $\beta_0 = \bar{\beta_0}$ and using the set of scaled instantaneous periodic waveguides with unit period [$\Lambda(z) = 1$].

\bibliography{bibliography}% Produces the bibliography via BibTeX.

\end{document}